\documentclass[submission,copyright,creativecommons]{eptcs}
\providecommand{\event}{PLACES 2024} 
\usepackage{iftex}

\ifpdf
  \usepackage{underscore}         
  \usepackage[T1]{fontenc}        
\else
  \usepackage{breakurl}           
\fi

\usepackage{amsmath}

\usepackage{amssymb}

\usepackage{amsthm}

\usepackage{etoolbox}

\usepackage{fontawesome}

\usepackage{graphicx}

\usepackage{hyperref}

\usepackage{listings}

\usepackage{mathpartir}

\usepackage{mathtools}

\usepackage{stmaryrd}

\usepackage{thmtools, thm-restate}

\usepackage{tikz}
\usetikzlibrary{decorations.pathreplacing,arrows.meta}

\usepackage[colorinlistoftodos]{todonotes}

\usepackage{wasysym}

\usepackage{xargs}

\usepackage{xcolor}

\newtoggle{meta-contents}
\toggletrue{meta-contents}

\newtoggle{todo-notes}
\toggletrue{todo-notes}

\newif\ifproof
\prooftrue

\newcommand{\keyword}[1]{\mathsf{#1}}
\newcommand{\letk}{\keyword{let}}
\newcommand{\boxk}{\keyword{box}}

\newcommand{\grammarEq}{::=}
\newcommand{\grammarOr}{\mid}

\newcommand{\substitution}[2]{[#1/#2]} 

\newcommand{\emptyseq}{\varepsilon}

\newcommand{\sendOp}{{!}}

\newcommand{\internalChoiceOp}{{\ensuremath{\oplus}}}
\newcommand{\externalChoiceOp}{{\ensuremath{\&}}}

\newcommand{\Unit}{*}

\newcommand{\Wait}{{\keyword{Wait}}}
\newcommand{\Close}{{\keyword{Close}}}

\newcommand{\modeType}[3]{{\ensuremath{#1 #2 . #3}}}
\newcommand{\choiceType}[3]{{\ensuremath{#1 \{#2\}^{#3}}}}

\newcommand{\functionType}[2]{{\ensuremath{#1 \rightarrow #2}}}
\newcommand{\productType}[2]{{\ensuremath{#1 \times #2}}}

\newcommand{\contextualType}[2]{{\ensuremath{(#1 \vdash #2)}}}
\newcommand{\contextualModalType}[1]{{\square#1}}

\newcommand{\binding}[3]{#1 :^{#2} #3}

\newcommand{\ctxlocal}[2]{#1^{<#2}}
\newcommand{\ctxouter}[2]{#1^{\ge #2}}

\newcommand{\fv}[1]{\ensuremath{\operatorname{free}\,#1}}

\newcommand{\unit}{*}

\newcommand{\receive}{\keyword{receive}}
\newcommand{\send}{\keyword{send}}

\newcommand{\abstraction}[3]{{\ensuremath{\lambda #1 . #3}}}
\newcommand{\application}[2]{{\ensuremath{#1 #2}}}
\newcommand{\pair}[2]{{\ensuremath{(#1, #2)}}}
\newcommand{\pairDestructor}[4]{{\ensuremath{\letk\ (#1, #2) = #3\ \keyword{in}\ #4}}}

\newcommand{\new}[1]{{\ensuremath{\keyword{new}\ #1}}}
\newcommand{\match}[3]{{\ensuremath{\keyword{match}\ #1\ \keyword{with}\ \{#2\}^{#3}}}}

\newcommand{\cbox}[1]{\ensuremath{\boxk\,#1}} 
\newcommand{\letbox}[3]{\ensuremath{\letk~\boxk~#1=#2~\keyword{in}~#3}}
\newcommand{\letunit}[2]{\ensuremath{\letk~*=#1~\keyword{in}~#2}}
\newcommand{\appliedvar}[2]{#1[#2]}
\newcommand{\ctxval}[2]{#1.#2}

\newcommand{\isterm}[3]{\ensuremath{#1 \vdash #2 : #3}}

\newcommand{\rulename}[1]{\textsc{#1}}
\newcommand{\rulearrowi}{\rulename{$\rightarrow$i}}
\newcommand{\rulearrowe}{\rulename{$\rightarrow$e}}
\newcommand{\ruleboxi}{\rulename{$\square$i}}
\newcommand{\ruleboxe}{\rulename{$\square$e}}
\newcommand{\rulevar}{\rulename{var}}
\newcommand{\rulesubs}{\rulename{subs}}
\newcommand{\rulectxt}{\rulename{ctxt}}
\newcommand{\ruleuniti}{\rulename{$\Unit$i}}
\newcommand{\ruleunite}{\rulename{$\Unit$e}}

\newcommand{\redOp}{\longrightarrow}
\newcommand{\reduction}[2]{\ensuremath{#1 \redOp #2}}

\newcommand{\etal}{et al.}

\newcommandx{\verified}[2][1=]{\todo[linecolor={\noteColor{verified}},backgroundcolor=cyan,bordercolor=black,size=\footnotesize,#1]{Verified: #2}}

\newcommandx{\corrected}[2][1=]{\todo[linecolor=green,backgroundcolor=green,bordercolor=black,size=\footnotesize,#1]{Corrected: #2}}

\newcommandx{\improve}[2][1=]{\todo[linecolor=yellow,backgroundcolor=yellow,bordercolor=black,size=\footnotesize,#1]{Improve: #2}}

\newcommandx{\change}[2][1=]{\todo[linecolor=red,backgroundcolor=red,bordercolor=black,size=\footnotesize,#1]{Change: #2}}

\newcommandx{\question}[2][1=]{\todo[linecolor=pink,backgroundcolor=pink,bordercolor=black,size=\footnotesize,#1]{Question: #2}}

\newcommandx{\suggestion}[2][1=]{\todo[linecolor=yellow,backgroundcolor=yellow,bordercolor=black,size=\footnotesize,#1]{Suggestion: #2}}

\newtheorem{lemma}{Lemma}

\numberwithin{corollary}{section}
\numberwithin{convention}{section}
\numberwithin{definition}{section}
\numberwithin{example}{section}
\numberwithin{lemma}{section}
\numberwithin{notation}{section}
\numberwithin{proposition}{section}
\numberwithin{remark}{section}
\numberwithin{theorem}{section}

\definecolor{darkviolet}{rgb}{0.5,0,0.4}
\definecolor{darkgreen}{rgb}{0,0.4,0.2}
\definecolor{darkblue}{rgb}{0.1,0.1,0.9}
\definecolor{darkgrey}{rgb}{0.5,0.5,0.5}
\definecolor{lightblue}{rgb}{0.4,0.4,1}

\lstdefinestyle{eclipse}{
  breaklines=true,
  basicstyle=\sffamily\small,
  emphstyle=\color{red}\bfseries,
  keywordstyle=\color{darkviolet}\bfseries,
  commentstyle=\color{darkgreen},
  stringstyle=\color{darkblue},
  numberstyle=\color{darkgrey},
  emphstyle=\color{red},
  showstringspaces=false,
}

\tikzset{
arrow/.style={>={Stealth[inset=0pt,length=8pt,angle'=28,round]}}}

\usepackage{cleveref}

\title{Linear Contextual Metaprogramming and Session Types}
\author{Pedro Ângelo
\institute{LIACC, Faculdade de Ciências da Universidade do Porto, Portugal}
\institute{LASIGE, Faculdade de Ciências da Universidade de Lisboa, Portugal}
\email{pjangelo@ciencias.ulisboa.pt}
\and
Atsushi Igarashi
\institute{Kyoto University, Kyoto, Japan}
\email{igarashi@kuis.kyoto-u.ac.jp}
\and
Vasco T. Vasconcelos
\institute{LASIGE, Faculdade de Ciências da Universidade de Lisboa, Portugal}
\email{vmvasconcelos@ciencias.ulisboa.pt}
}

\def\titlerunning{Linear Contextual Metaprogramming and Session Types}
\def\authorrunning{P. Ângelo, A. Igarashi \& V. T. Vasconcelos}

\begin{document}

\maketitle

\begin{abstract}
  We explore the integration of metaprogramming in a call-by-value linear
  lambda-calculus and sketch its extension to a session type system.
  We build on a model of contextual modal type theory with multi-level contexts,
  where contextual values, closing arbitrary terms over a series of variables,
  may then be boxed and transmitted in messages. Once received, one such value
  may then be unboxed (with a let-box construct) and locally applied before
  being run.
  We present a series of examples where servers prepare and ship code on demand
  via session typed messages.

\end{abstract}


\section{Introduction}
\label{sec:intro}

Metaprogramming manipulates code in order to generate and evaluate code at
runtime, allowing in particular to explore the availability of certain arguments
to functions in order to save computational effort.
%
In this paper we are interested in programming languages where the code produced
is typed by construction and where code may refer to a context providing types
for the free variables, commonly known as contextual
typing~\cite{DBLP:journals/pacmpl/JangGMP22,DBLP:conf/esop/MuraseNI23,DBLP:journals/jfp/NanevskiP05,DBLP:journals/tocl/NanevskiPP08}.
On an orthogonal axis, session types have been advocated as a means to discipline
concurrent computations, by accurately describing protocols for the channels
used to exchange messages between processes~\cite{DBLP:journals/iandc/AlmeidaMTV22,DBLP:journals/jfp/GayV10,DBLP:conf/concur/Honda93,DBLP:conf/esop/HondaVK98,DBLP:conf/parle/TakeuchiHK94,DBLP:conf/icfp/ThiemannV16,DBLP:journals/iandc/Vasconcelos12}.

The integration of session types with metaprogramming allows to setup
code-producing servers that run in parallel with the rest of the program and
provide code on demand, exchanged on typed channels.
Linearity is central to session types, but current metaprogramming models lack
support for such a feature. We extend a simple model of contextual modal type
theory (with monomorphic contexts) with support for linear types, to obtain a
call-by-value linear lambda calculus with multi-level contexts. We then sketch
how to extend this language with support for concurrency and session types.

Our development is based on Davies and
Pfenning~\cite{DBLP:journals/jacm/DaviesP01}, where we use a box modality to
distinguish generated code. We further allow code to refer to variables in a
context, described by contextual types, along the lines of Nanevski
\etal~\cite{DBLP:journals/tocl/NanevskiPP08}. M\oe
bius~\cite{DBLP:journals/pacmpl/JangGMP22} further adds to modal contextual type
theory the provision for pattern matching on code, for generating polymorphic
code, and for generating code that depends on other code fragments. We forgo the
first two directions, and base our development on the last.
We propose a multilevel contextual modal \emph{linear} lambda calculus, where in
particular the composition of code fragments avoids creating extraneous
administrative redexes due to boxing and unboxing.

An alternative starting point would have been the Fitch- or Kripke-style
formulation, providing for the Lisp quote/unquote, where typing contexts are
viewed as stacks modeling the different stages of
computation~\cite{DBLP:conf/esop/MuraseNI23}. It seemed to us that the let-box
approach would simplify the extension to the linear setting, and then to session
types.


To motivate the problem let us start with the issue of generating code to
send a fixed number of integers on a stream. The type of streams, as seen from
the side of processes writing on the stream is as follows.
\begin{lstlisting}
type Stream = oplus{More: !Int.Stream, Done: Close}
\end{lstlisting}
The writer chooses between selecting \lstinline|More| values or selecting
\lstinline|Done|. In the former case, the writer sends an integer value and
``goes back to the beginning''; in the latter case the writer must close the
channel.
Function \lstinline|sendFives| accepts an integer value and returns a code
fragment that requires a \lstinline|Stream| and, when executed, produces a unit
value, written \lstinline|[Stream vdash Unit]|. We proceed by pattern-matching
on the parameter.
\begin{lstlisting}
sendFives : Int -> [StreamvdashUnit]
sendFives 0 = box (y. close (select Done y))
sendFives n = let box u = sendFives (n - 1)
              in box (x. u[send 5 (select More x)])
\end{lstlisting}
When all values have been sent in the stream (when \lstinline|n| is
\lstinline|0|), all it remains is to
\lstinline|select Done| and then close the channel. The \lstinline|box|
expression generates code under a variable environment (an evaluation context),
in this case containing variable \lstinline|y| alone, the channel endpoint.
Otherwise, we recursively compute code to send \lstinline|n-1| values, unbox it
storing the result in \lstinline|u|, and then prepare code to send the
\lstinline|n|-th value. Expression \lstinline|u[send 5 (select More x)]| (an
applied variable) applies expression \lstinline|send 5 (select More x)| of type
\lstinline|Stream|to \lstinline|u| (a contextual value of type
\lstinline|(Stream vdash Unit|) to obtain an expression of type \lstinline|Unit|.
If \lstinline|u| is the contextual value \lstinline|y. close (select Done y)|,
then the explicit substitution evaluates to \lstinline|close (select Done send 5 (select More x))|.

We may now compute and run code to send a fixed number of integer values.
\begin{lstlisting}
send4Fives : Stream -> Unit
send4Fives c = let box u = sendFives 4 in u[c]
\end{lstlisting}
%
Expression \lstinline|sendFives 4| is a boxed code fragment of type
\lstinline|[Stream vdash Unit]|. Then, \lstinline|u| is an (unboxed) code
fragment of contextual type \lstinline|(Stream vdash Unit)|. We provide the code
fragment with an explicit substitution \lstinline|[c]|. The whole \lstinline|let|
expression then amounts to running the code
\begin{lstlisting}
  close (select Done (send 5 (select More (...send 5 (select More c)...))))
\end{lstlisting}
without calling function \lstinline|sendFives| or using recursion in any other form.







The next example transmits code on channels. Imagine a server preparing code on
behalf of clients. The server uses a channel to interact with its clients: it
first receives a number \lstinline|n|, then replies with code to send
\lstinline|n| fives, and finally waits for the channel to be closed.
The type of the communication channel is as follows.
\begin{lstlisting}
type Builder = ?Int.![StreamvdashUnit].Wait
\end{lstlisting}

The server receives \lstinline|n| on a given channel and computes the code using
a call to \lstinline|sendFives|:
\begin{lstlisting}
serveFives : Builder -> Unit
serveFives c =
  let (n,c) = receive c in wait (send (sendFives n) c)
\end{lstlisting}

On the other end of the channel sits a client: it sends a number (4 in this
case), receives the code (of type \lstinline|[Stream vdash Unit]|), closes the
channel and evaluates the code received.
\begin{lstlisting}
sendFives' : Dual Builder -> Stream -> Unit
sendFives' c d =
  let (code,c) = receive (send 4 c) in close c ;
  let box u = code in u[d]
\end{lstlisting}
The \lstinline|Dual| operator on session types provides a view of the other end
of the channel. In this case, \lstinline|Dual Builder| is the type
\lstinline|!Int.?[Stream vdash Unit].Close|, where \lstinline|!| is turned into
\lstinline|?| and \lstinline|Wait| is turned into
\lstinline|Close| (and conversely in both cases).
Notice that \lstinline|code| is a boxed code fragment of type 
\lstinline|[Stream vdash Unit]|, hence \lstinline|u| is the corresponding code
fragment (of type \lstinline|(Stream vdash Unit)|) and \lstinline|u[d]| runs the
code on channel \lstinline|d|.

To complete the example we need a function for reading streams, a consumer of
type \lstinline|Dual Stream -> Unit|. Function \lstinline|readInts| reads and
discards all integer values on the stream and then waits for the stream to be
closed.
\begin{lstlisting}
readInts : Dual Stream -> Unit
readInts (Done c) = wait c
readInts (More c) = let (_, c) = receive c in readInts c
\end{lstlisting}

Finally, the main thread forks two threads---one running \lstinline|serveFives|,
the other to collect the integer values (\lstinline|readInts|)---and continues with
\lstinline|sendFives'|. We take advantage of a primitive function,
\lstinline|forkWith| that expects a suspended computation (a thunk), creates a
new channel, forks the thunk on one end of the channel, and returns the other
end of the channel for further interaction.
\begin{lstlisting}
main : Unit
main =
  let c = forkWith (lambda_->serveFives) in -- c : Dual Builder
  let d = forkWith (lambda_->readInts) in   -- d : Stream
  sendFives' c d
\end{lstlisting}

The interaction among the three processes is depicted as follows

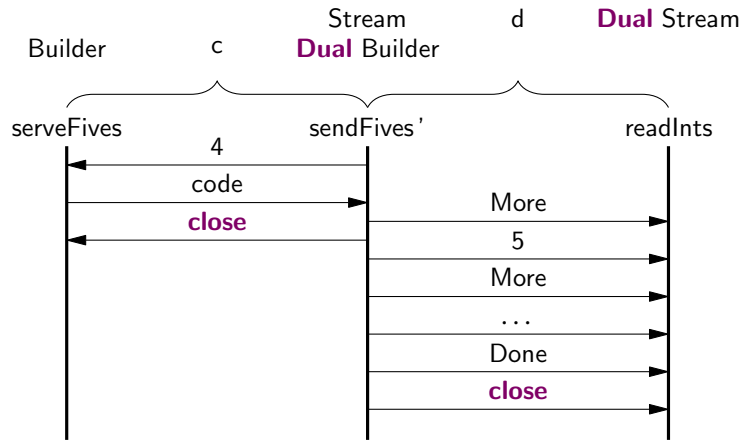
\begin{center}
\begin{tikzpicture}
\node (process1Begin) [align=center] {\lstinline|serveFives|};
\node (process1End) [below of=process1Begin, node distance=4.3cm] {};
\draw [draw, very thick, -] (process1Begin.south) -- (process1End.north);

\node (process2Begin) [align=center, right of=process1Begin, node distance=4cm] {\lstinline|sendFives'|};
\node (process2End) [below of=process2Begin, node distance=4.3cm] {};
\draw [draw, very thick, -] (process2Begin.south) -- (process2End.north);

\node (process3Begin) [align=center, right of=process2Begin, node distance=4cm] {\lstinline|readInts|};
\node (process3End) [below of=process3Begin, node distance=4.3cm] {};
\draw [draw, very thick, -] (process3Begin.south) -- (process3End.north);

\draw [decorate,decoration={brace,amplitude=10pt}]  (process1Begin.north) -- (process2Begin.north) node[near start, xshift=-1cm, yshift=0.8cm]{\lstinline|Builder|} node[midway,yshift=0.8cm]{\lstinline|c|} node[near end, xshift=1cm, yshift=0.8cm]{\lstinline|Dual Builder|};

\draw [decorate,decoration={brace,amplitude=10pt}]  (process2Begin.north) -- (process3Begin.north) node[near start, xshift=-1cm, yshift=1.2cm]{\lstinline|Stream|} node[midway,yshift=1.2cm]{\lstinline|d|} node[near end, xshift=1cm, yshift=1.2cm]{\lstinline|Dual Stream|};

\draw [draw, <-, arrow] ([yshift=-.25cm]process1Begin.south) -- node [above] {\lstinline|4|} ([yshift=-.25cm]process2Begin.south);
\draw [draw, ->, arrow] ([yshift=-.75cm]process1Begin.south) -- node [above] {\lstinline|code|} ([yshift=-.75cm]process2Begin.south);
\draw [draw, <-, arrow] ([yshift=-1.25cm]process1Begin.south) -- node [above] {\lstinline|close|} ([yshift=-1.25cm]process2Begin.south);

\draw [draw, ->, arrow] ([yshift=-1cm]process2Begin.south) -- node [above] {\lstinline|More|} ([yshift=-1cm]process3Begin.south);
\draw [draw, ->, arrow] ([yshift=-1.5cm]process2Begin.south) -- node [above] {\lstinline|5|} ([yshift=-1.5cm]process3Begin.south);
\draw [draw, ->, arrow]  ([yshift=-2cm]process2Begin.south) -- node [above] {\lstinline|More|} ([yshift=-2cm]process3Begin.south);
\draw [draw, ->, arrow] ([yshift=-2.5cm]process2Begin.south) -- node [above] {\dots} ([yshift=-2.5cm]process3Begin.south);
\draw [draw, ->, arrow] ([yshift=-3cm]process2Begin.south) -- node [above] {\lstinline|Done|} ([yshift=-3cm]process3Begin.south);
\draw [draw, ->, arrow] ([yshift=-3.5cm]process2Begin.south) -- node [above] {\lstinline{close}} ([yshift=-3.5cm]process3Begin.south);
\end{tikzpicture}
\end{center}
where the \lstinline|code| produced by function \lstinline|serveFives| and
transmitted to \lstinline|sendFives'| is
\begin{lstlisting}
  box (close (select Done (send 5 (select More (...(select More c)...)))))
\end{lstlisting}

In the rest of the paper we develop our system. In \cref{section:linear-staged}
we introduce the call-by-value linear lambda calculus with multi-level contexts
and in \cref{sec:session-staged} we sketch the extensions required to type and
run the examples in this section. We conclude in \cref{section:conclusion}.


\section{Linear staged metaprogramming}
\label{section:linear-staged}

This section introduces the call-by-value linear lambda calculus with
multi-level contexts.

\paragraph{Syntax}

Our language is defined over the syntactic categories of set of variables,
$x,y,z$.
We write $\overline X$ for a sequence of objects $X_1\cdots X_n$ with $n\ge0$.
The empty sequence (when $n=0$) is denoted by~$\emptyseq$.
  \begin{alignat*}{4}
    &\text{Contextual type} \quad &&\tau\quad && \grammarEq \quad &&
    \contextualType{\overline \tau}T
    \\
    &\text{Type} \quad &&T, U\quad && \grammarEq \quad && \Unit
    \grammarOr \functionType{T}{T} 
    \grammarOr \contextualModalType \tau
    \\
    &\text{Contextual value} \quad &&\rho,\sigma \quad && \grammarEq \quad &&
    \ctxval{\overline x}M
    \\
    &\text{Value} \quad &&v \quad && \grammarEq \quad &&
    \unit
    \grammarOr \abstraction{x}{T}{M}
    \grammarOr \cbox \sigma
    \\
    &\text{Term} \quad &&M, N \quad && \grammarEq \quad &&
    v
    \grammarOr \appliedvar x{\overline \sigma}
    \grammarOr \letunit MM
    \grammarOr \application{M}{M}
    \grammarOr \letbox xMM
    \\
    &\text{Level} \quad &&k,m,n \quad && \in \mathbb{N} &&
    \\
    &\text{Typing context} \quad &&\Gamma, \Delta \quad && \grammarEq \quad &&
    \emptyseq \grammarOr \Gamma, \binding{x}{n}{\tau}
  \end{alignat*}

Types include the unit ($\Unit$) linear type, the linear arrow type
$\functionType{T}{U}$, and the linear boxed contextual type
$\contextualModalType\tau$. A contextual type $\tau$ of the form
$\contextualType{\overline\tau}{T}$ represents code of type $T$, parameterized on
variables typed by a list of contextual types $\overline\tau$, called
$\emph{contexts}$~\cite{DBLP:journals/pacmpl/JangGMP22,DBLP:conf/esop/MuraseNI23}.


To objects of the form $\ctxval{\overline x}M$ we call \emph{contextual values}.
They denote code fragments $M$ parameterized by the variables in sequence
$\overline x$.
The values in the language include $\unit$ (introducing type $\Unit$),
lambda abstraction (introducing the arrow type), and the $\keyword{box}$ term
(introducing the contextual modal type $\contextualModalType\tau$).
Terms include values, contextual term variables
$\appliedvar{x}{\overline\sigma}$ applying contextual values $\overline{\sigma}$
to the code fragment described by $x$, the $\keyword{let}~\unit$ (eliminating
$\unit$), lambda application (eliminating the arrow type), and the
$\keyword{let~box}$ term (eliminating the contextual modal type
$\contextualModalType\tau$).


When $\overline x$ is the empty sequence we sometimes write $M$ in place of the
contextual value $\ctxval \emptyseq M$. Similarly, when $\overline\sigma$ is the
empty sequence we sometimes write $x$ instead of the contextual term variable
$\appliedvar{x}{\emptyseq}$.
Furthermore, in examples we write $\keyword{Unit}$ in place of $\Unit$, and
$[\overline \tau \vdash T]$ in place of
$\contextualModalType{\contextualType{\overline \tau}{T}}$.


The bindings in the language are the following. Variable $x$ is bound in 
$M$ (but not in $N$) in terms $\abstraction{x}{}{M}$
and $\letbox xNM$.
The set of bound and free variables in terms ($\fv M$) are defined accordingly.
We follow the variable convention whereby terms that differ only in the names
of the bound variables are interchangeable in all
contexts~\cite{DBLP:books/daglib/0005958}.


\paragraph{Contexts for local and outer variables}



Typing contexts bind contextual types to variables; we annotate each entry with
its level $n$.
We assume that contexts contain no duplicate variables and write $\Gamma,\Delta$
for the context containing the entries in both $\Gamma$ and $\Delta$, provided
they feature disjoint sets of variables.

We introduce two predicates on typing contexts. If
$\Gamma = \binding{x_1}{k_1}{\tau_1}, \dots, \binding{x_m}{k_m}{\tau_m}$, then
$\ctxlocal \Gamma n$ holds when $n$ is greater than all the levels of the
bindings in $\Gamma$ (that is, $k_1$, \dots, $k_m$), and $\ctxouter \Gamma n$
holds when $n$ is smaller or equal than all the levels in $\Gamma$. More precisely,
\begin{align*}
  \ctxlocal \Gamma n \text{ holds when } \max(0, k_1,\dots k_m) < n
  &\qquad&
  \ctxouter \Gamma n \text{ holds when } m = 0 \text{ or } \min(k_1,\dots k_m) \ge n
\end{align*}
%
%
Intuitively, $\ctxlocal \Gamma n$ denotes the local variables in a code fragment
of level $n$, whereas $\ctxouter \Gamma n$ denotes the outer variables. Notice
that there is no context $\Gamma$ for which $\ctxlocal \Gamma 0$ (not even the
empty context), meaning that code fragments start at level 1.

\paragraph{Substitution}


Substitution is that of the linear lambda calculus except for the fact that we
use applied modal variables $\appliedvar x{\overline \sigma}$ rather than
ordinary variables $x$.
We denote by $\substitution{\sigma}{x}M$ the term obtained by replacing variable
$x$ by the contextual value $\sigma$ in term $M$.
We detail some illustrative cases.
\begin{align*}
  \substitution{\ctxval{\overline z}M}{x}(\appliedvar x{\overline\rho})
  & = \substitution{\overline\rho}{\overline z} M
  \\
  \substitution \sigma x (\letbox yMN) & =
  \begin{cases}
    \letbox y{\substitution \sigma xM}N & \text{if } x\in\fv M
    \\
    \letbox y M{\substitution \sigma xN} & \text{if } x\in\fv N
  \end{cases}
\end{align*}
Substitution on $\letk~\unit$ and on application is similar to $\letk~\boxk$.
Substitution on the remaining type constructors is an homomorphism; for example
$\substitution{\sigma}x(\cbox{(\ctxval{\overline y}{M})}) =
\cbox{(\ctxval{\overline y}{\substitution{\sigma}xM})}$. Substitution on applied
variables,
$\substitution{\ctxval{\overline z}{M}}{x}(\appliedvar{y}{\overline\rho})$, is
defined only when $x$ and $y$ coincide and when $\overline z$ and
$\overline\rho$ are sequences of the same length (the variable convention
ensures that the variables in $\overline z$ are pairwise distinct and not free
outside $M$, hence the substitution of the various variables in $\overline z$
can be performed in sequence). Substitution in $\letk~\boxk$ is defined only
when $x\in\fv(\application MN)$; it is undefined when $x$ is both free in $M$
and in $N$. The typing system guarantees substitution is defined for all typable
processes.
Typing judgments are of the form $\isterm \Gamma M T$ stating that term $M$ has
type $T$ under typing context $\Gamma$; the rules are introduced below.
\begin{lemma}[Substitution principle]
  \begin{mathpar}
  \inferrule*[right=\rulesubs]{
    \isterm{\Gamma,\binding xn{\tau}}{M}{T}
    \\
    \isterm{\ctxouter{\Delta}{n}}{\sigma}{\tau}
  }{
    \isterm{\Gamma, \ctxouter{\Delta}{n}}{\substitution{\sigma}{x}M}{T}
  }    
\end{mathpar}
\end{lemma}
%

When the contextual term variable denotes a parameterless code fragment, we
recover the conventional substitution principle in linear type systems. If we
write $\contextualType\emptyseq T$ simply as $T$, and the contextual value
$\ctxval\emptyseq N$ as $N$, we have
\begin{mathpar}
  \inferrule{
    \isterm{\Gamma,\binding xn{\contextualType\emptyseq U}}{M}{T}
    \\
    \isterm{\ctxouter{\Delta}n}{\ctxval\emptyseq N}{\contextualType\emptyseq U}
  }{
    \isterm{\Gamma, \ctxouter{\Delta}n}{\substitution {\ctxval\emptyseq N}xM}{T}
  }    

  \text{ abbreviated to}
  
  \inferrule{
    \isterm{\Gamma,\binding x{}{U}}{M}{T}
    \\
    \isterm{\Delta}{N}{U}
  }{
    \isterm{\Gamma,\Delta}{\substitution NxM}{T}
  }    
\end{mathpar}
%


\paragraph{Evaluation}

Evaluation on terms is given by the relation $\reduction{M}{N}$ defined by the
axioms
\begin{align*}
  \reduction{\application{(\abstraction{x}{T}{M})}{v} &}{\substitution{\ctxval{\emptyseq}v}{x}M}
  \\
  \reduction{\letunit \unit M &}{M}
  \\
  \reduction{\letbox x {\cbox\sigma} M &}{\substitution{\sigma}{x}}M
\end{align*}
compounded by the usual congruence rules of the call-by-value
$\lambda$-calculus. We do not allow reduction inside a box term (a value) as we
do not allow reduction under a $\lambda$ (another value).


\paragraph{Typing contextual terms}

Contextual terms are closures of the form $\ctxval{\overline x}{M}$, closing
term $M$ over a sequence of variables $\overline x$. Such a contextual term is
given a contextual type $\contextualType{\overline \tau}T$ when $M$ is of type $T$
under the hypothesis that the variables in $\overline x$ are of the types in $\overline \tau$.
%
\begin{mathpar}
  \inferrule*[right=\rulectxt]{
    \isterm{\ctxouter{\Gamma}{n},\ctxlocal{\overline{\binding{x}{k}{\tau}}}{n}}{M}{T}
  }{
    \isterm{\ctxouter{\Gamma}{n}}{\ctxval{\overline x}{M}}{\contextualType{\overline \tau}{T}}
  }
\end{mathpar}
Intuitively, the free variables of $M$ are split in two groups: local and outer.
The local variables, $\overline x$, are distinguished in the contextual term
$\ctxval{\overline x}M$ and typed under context
$\ctxlocal{\overline{\binding xk\tau}} n$ where $n$ is an upper bound of the levels
$k_i$ assigned to each local variable. The level of each local variable is
arbitrary, as long as it is lower than $n$.
Outer variables are typed under context $\ctxouter \Gamma n$.
Natural number $n$ thus denotes the level at which the code fragment
$\ctxval{\overline x}M$ is typed. In general, a contextual term $\sigma$ can be
typed at different levels. All resources (variables) in the context $\Gamma$ in
the conclusion are consumed in the premise; furthermore resources $\overline x$
are consumed in the derivation of $M$ (implying that they must be free in $M$).


\paragraph{Typing variables}

Variables are of the form $\appliedvar{x}{\sigma_1\cdots \sigma_m}$, denoting a
contextual term applied to contextual values $\sigma_1\cdots \sigma_m$. To type
each $\sigma_i$ we need a separate context which we call $\Gamma_i$. The type of
$x$ must be a contextual type of the form
$\contextualType{\tau_1\cdots \tau_m}{T}$ and each contextual term $\sigma_i$
must be of type $\tau_i$.
%
We can easily see that all resources in the typing context in the
conclusion are consumed: the various $\Gamma_i$ are consumed in the premises,
that for $x$ is consumed at the right of the turnstile. This guarantees that all
resources are used and thus that the rule includes no implicit form of weakening.
\begin{mathpar}
  \inferrule*[right=\rulevar]{
    \isterm{\overline\Gamma}{\overline\sigma}{\overline\tau}
  }{
    \isterm{\overline{\Gamma}, \binding{x}{n}{
        \contextualType{\overline{\tau}}{T}}}{\appliedvar{x}{\overline\sigma}}{T}
  }    
\end{mathpar}
The lengths of all sequences, $\overline{\Gamma}$, $\overline{\sigma}$ and
$\overline{\tau}$, must coincide. When the sequences are empty, the rule becomes
an axiom. In fact rule \rulevar{} is the natural generalization of the axiom in
the linear $\lambda$-calculus, obtained when writing type
$\contextualType{\emptyseq}{T}$ in the context as $T$, writing applied variable
$\appliedvar x \emptyseq$ as $x$, and omitting the level of the variable in the
context.

\begin{mathpar}
  \inferrule{
  }{
    \isterm{\binding{x}{n}{\contextualType{\emptyseq}{T}}}{\appliedvar
        x \emptyseq}{T}
  }    

  \text{abbreviated to}

  \inferrule{
  }{
    \isterm{\binding{x}{}{T}}{x}{T}
  }    
\end{mathpar}

\paragraph{Typing $\lambda$ abstraction and application}

We now address the conventional typing rules for the introduction and
elimination of the arrow in the linear lambda calculus; the rules are below.
\begin{mathpar}
  \inferrule*[right=\rulearrowi]{
    \isterm{\Gamma,\binding{x}{0}{\contextualType{\emptyseq}{T}}}{M}{U}
  }{
    \isterm{\Gamma}{\abstraction{x}{}{M}}{\functionType{T}{U}}
  }

  \inferrule*[right=\rulearrowe]{
    \isterm{\Gamma}{M}{\functionType{T}{U}}
    \\
    \isterm{\Delta}{N}{T}
  }{
    \isterm{\Gamma,\Delta}{\application MN}{U}
  }  
\end{mathpar}
The $\lambda$-bound variable $x$ is local to $M$, hence of level $0$.

\emph{Local soundness}. Notice that $\ctxouter\Delta 0$ is true of all contexts $\Delta$.
\begin{mathpar}
  \inferrule*[right=\rulearrowe]{
    \inferrule*[right=\rulearrowi]{
      \isterm{\Gamma,\binding x0{\contextualType{\emptyseq}{T}}} M U
    }{
      \isterm \Gamma {\abstraction x{}M}{\functionType TU}
    }
    \;\;\; 
    \isterm \Delta v T
  }{
    \isterm{\Gamma,\Delta}{\application{(\abstraction x{}M)}{v}}{U}
  }
  \;\Rightarrow\;
  \inferrule*[right=\rulesubs]{
    \isterm{\Gamma,\binding x0{\contextualType{\emptyseq}{T}}}{M}{U}
    \quad 
    \inferrule*[right=\rulectxt]{
      \isterm{\Delta}{v}{T}
    }{
      \isterm{\Delta}{\ctxval \emptyseq v}{\contextualType \emptyseq T}
    }
  }{
    \isterm{\Gamma,\Delta}{\substitution {\ctxval{\emptyseq}{v}}xM}{U}
  }   
\end{mathpar}

\emph{Local completeness}:
\begin{mathpar}
  \isterm{\Gamma,\Delta} M {\functionType TU}
  \quad\Rightarrow\quad
  \inferrule*[right=\rulearrowi]{
    \inferrule*[right=\rulearrowe]{
      \isterm{\Gamma,\Delta} M {\functionType TU}
      \\
      \inferrule*[right=\rulevar]{ }{\isterm{\binding
          x0{\contextualType{\emptyseq}{T}}} {\appliedvar x \emptyseq} T}
    }{
      \isterm{\Gamma,\Delta,\binding x0{\contextualType \emptyseq T}} {\application M{(\appliedvar x \emptyseq)}} U
    }
  }{
      \isterm{\Gamma,\Delta} {\abstraction x {} {\application M{(\appliedvar x \emptyseq)}}} {\functionType TU}
  }
\end{mathpar}

\paragraph{Typing \Unit{} introduction and elimination}
Rules:
\begin{mathpar}
  \inferrule*[right=\ruleuniti]{
  }{
    \isterm{}{\unit}{\Unit}
  }

  \inferrule*[right=\ruleunite]{
    \isterm{\Gamma}{M}{\Unit}
    \\
    \isterm{\Delta}{N}{T}
  }{
    \isterm{\Gamma,\Delta}{\letunit{M}{N}}{T}
  }
\end{mathpar}

\emph{Local soundness and local completeness}:
\begin{mathpar}
  \inferrule*[right=\ruleunite]{
    \inferrule*[right=\ruleuniti]{ }{
      \isterm{}{\unit}{\Unit}
      }
      \\
      \isterm \Gamma M T
  }{
    \isterm{}{\letunit{\unit}{M}}{T}
  }
  \quad\Leftrightarrow\quad
  \isterm \Gamma M T
\end{mathpar}

\paragraph{Typing box introduction and elimination}

The box introduction and elimination rules are as follows.
\begin{mathpar}
  \inferrule*[right=\ruleboxi]{
    \isterm{\Gamma} \sigma \tau
  }{
    \isterm{\Gamma} {\cbox\sigma}{\contextualModalType{\tau}}
  }
  
  \inferrule*[right=\ruleboxe]{
    \isterm{\ctxouter \Gamma n} {M}{\contextualModalType\tau}
    \\
    \isterm{\Delta,\binding x n {\tau}} {N}{T}
  }{
    \isterm{\ctxouter \Gamma n,\Delta}{\letbox{x}{M}{N}}{T}
  }
\end{mathpar}
In rule \ruleboxi, the type of the box is the contextual modal type
$\contextualModalType\tau$ if the contextual term $\sigma$ has type $\tau$.
The context for the box elimination rule, \ruleboxe{}, is split into two: one
part ($\ctxouter \Gamma n$) to type $M$, the other ($\Delta$) to type $N$. Term
$M$ must denote a code fragment, hence the type of $M$ must be a contextual
modal type $\contextualModalType\tau$. Term $N$ is typed under
context $\Delta$ extended with an entry for $x$.
The context in the conclusion is formed by the juxtaposition of the contexts in
the premises thus ensuring that all resources are fully consumed.
The level $n$ of variable $x$ determines the level of code $M$.
In contrast to preceding
work~\cite{DBLP:journals/jacm/DaviesP01,DBLP:journals/pacmpl/JangGMP22,DBLP:journals/jfp/NanevskiP05},
level control is shifted from box introduction to the rule that types contextual
values. A rule derived from \rulectxt{} followed by \ruleboxi{} coincides with
the box introduction rule of M\oe bius~\cite{DBLP:journals/pacmpl/JangGMP22}:
\begin{mathpar}
  \inferrule{
    \isterm{\ctxouter{\Gamma}{n},\ctxlocal{\overline{\binding{x}{k}{\tau}}}{n}}{M}{T}
  }{
    \isterm{\ctxouter{\Gamma}{n}}{\cbox{(\ctxval{\overline x}{M})}}{\contextualModalType{\contextualType{\overline\tau}{T}}}
  }
\end{mathpar}

\emph{Local soundness}:
\begin{mathpar}
  \inferrule*[right=\ruleboxe]{
    \inferrule*[right=\ruleboxi]{
      \isterm{\ctxouter \Gamma n} \sigma \tau
    }{
      \isterm{\ctxouter \Gamma n} {\cbox\sigma}{\contextualModalType{\tau}{}}
    }
    \\
    \isterm{\Delta,\binding x n \tau} {M}{T}
  }{
    \isterm{\ctxouter \Gamma n,\Delta}{\letbox{x}{\cbox\sigma}{M}}{T}
  }
  \quad\Rightarrow\quad
  \inferrule*[right=\rulesubs]{
    \isterm{\Delta,\binding xn{\tau}}{M}{T}
    \\
    \isterm{\ctxouter{\Gamma}{n}}{\sigma}{\tau}
  }{
    \isterm{\ctxouter \Gamma n,\Delta}{\substitution{\sigma}{x}M}{T}
  }    
\end{mathpar}

\emph{Local completeness}:
\begin{mathpar}
  \isterm {\ctxouter\Gamma2} M {\contextualModalType\tau}
  \quad\Rightarrow\quad
  \inferrule*[right=\ruleboxe]{
    \isterm {\ctxouter\Gamma2} M {\contextualModalType\tau}
    \\
    \inferrule*[right=\ruleboxi]{
      \inferrule*[right=\rulectxt]{
        \inferrule*[right=\rulevar]{
          \inferrule*[right=\rulectxt]{
            \inferrule*[right=\rulevar]{
            }{
              \isterm{\binding {x_i}{1}{\contextualType\emptyseq{U_i}}}{\appliedvar{x_i}\emptyseq}{U_i}
            }
          }{
            \isterm{\binding{x_i}{1}{\contextualType{\emptyseq}{U_i}}}{\ctxval \emptyseq{\appliedvar{x_i}{\emptyseq}}}{\contextualType\emptyseq{U_i}}
          }
        }{
          \isterm {\binding u 2 {\tau}, \overline{\binding{x}{1}{\contextualType{\emptyseq}{U}}}} {\appliedvar{u}{\overline{\ctxval\emptyseq{\appliedvar x\emptyseq}}}} {T}
        }
      }{
        \isterm{\binding u 2 {\tau}} {\ctxval{\overline x}{\appliedvar{u}{\overline{\ctxval\emptyseq{\appliedvar x\emptyseq}}}}} {\tau}
      }
    }{
      \isterm{\binding u 2 {\tau}}{\cbox{(\ctxval{\overline x}{\appliedvar{u}{\overline{\ctxval\emptyseq{\appliedvar x\emptyseq}}}})}}{\contextualModalType\tau}
    }
  }{
    \isterm{\ctxouter\Gamma2}{\letbox{u}{M}{\cbox{(\ctxval{\overline x}{\appliedvar{u}{\overline{\ctxval\emptyseq{\appliedvar x\emptyseq}}}})}}}{\contextualModalType\tau}
  }
\end{mathpar}
where $\tau$ abbreviates
$\contextualType {\overline{\contextualType{\emptyseq}{U}}} T$ and $\overline U
= U_1 \cdots U_m$ and similarly for $\overline x$.

\paragraph{An example from M\oe bius}

Concrete syntax apart, the $\beta$-reduction for box terms is that of
M{\oe}bius~\cite{DBLP:journals/pacmpl/JangGMP22}. We adapt an example to make it
linear.
\begin{lstlisting}
let box r = box (y. y + 2) in
let box u = box (c,x. 3 * z + c[2 * x]) in
            box (y. u[y. r[y], y])
\end{lstlisting}
reduces to
\begin{lstlisting}
let box u = box (c,x. 3 * z + c[2 * x]) in
            box (y. u[y. y + 2, y])
\end{lstlisting}
under substitution \lstinline|[(y. y + 2) / r]|, which in turn reduces to
\begin{lstlisting}
box (y. 3 * z + 2 * y + 2)
\end{lstlisting}
under substitution \lstinline|[(c,x. 3 * z + c[2 * x]) / u]|, generating no
administrative redexes.
The original redex (or any contractum) has type \lstinline|boxt (vdashInt)|
under typing context \lstinline|z :^3 (vdashInt)|, assuming suitable rules for
integer constants and arithmetic operators.
Contextual value \lstinline|(y. y + 2)|, and hence also the contextual variable
\lstinline|r|, is typed at level \lstinline|2|.
Contextual value \lstinline|(c,x. 3 * z + c[2 * x])|, and hence also contextual
variable \lstinline|u|, is typed at level \lstinline|2|.
Variable \lstinline|z| can be typed at level \lstinline|2|.


\section{From linear staged to session staged metaprogramming}
\label{sec:session-staged}

This section briefly outlines what it takes to bridge the gap between the linear
lambda calculus with multi-level contexts and the language required to type and
run the examples in the introduction.

First and foremost, \emph{session types} must be introduced in the syntax of
types. In the absence of polymorphism, a new syntactic category $S$ may be
introduced for session types. Session types include input and output ($?T.S$ and
$!T.S$), branch and select ($\choiceType{\externalChoiceOp}{l: S_l}{l \in L}$
and $\choiceType{\internalChoiceOp}{l: S_l}{l \in L}$), and channel closing
($\Wait$ and $\Close$), where $l$ denotes a label and $L$ a label set. In
addition, and in order to express the \lstinline|Stream| type in the example, we
need recursive types. They are usually introduced in the form of a $\mu$ (or
$\keyword{rec}$) constructor and type references (sometimes called type
variables), and treated equi-recursively.
Session types then become an extra constructor for types
$T$~\cite{DBLP:conf/esop/HondaVK98,DBLP:journals/iandc/Vasconcelos12}.
For a more ambitious setting one may consider the sequential composition of
types ($R;S$) and continuation-less input and output ($?T$ and $!T$) as in
context-free session
types~\cite{DBLP:journals/iandc/AlmeidaMTV22,DBLP:conf/icfp/ThiemannV16}.

At the level of \emph{terms}, we require a few session-related primitives. We
need constants to receive, to send, to select a choice, to wait for a channel to
be closed and to close a channel. These can be given type schemes. For example,
the type of constant $\send$ can be given by a type of the form
$\functionType{T}{\functionType{\modeType{\sendOp}{T}{S}}{S}}$, a function
receiving a value to be sent and a channel on which to send the value, and
returning the channel on which to continue interaction.
We further need a constant to fork a new thread whose type can be given by type
$\functionType{(\functionType{\Unit}{\Unit})}{\Unit}$, accepting a thunk and
giving back a unit vale. Branching (achieved by pattern-matching in the
examples) cannot be given by a constant. We then add a new term constructor
$\match{M}{l\colon M_l}{l \in L} $. Finally we need a primitive to create a new
channel, usually of the form $\new{S}$, returning the two end points of the
newly created channel.
Both the primitive $\receive$ operation and channel creation $\new{S}$ return a
pair; we then need \emph{linear} pairs of type $\productType TU$, with
introduction $\pair MN$ and elimination $\pairDestructor xyMN$ terms.
Details can be found in Gay and Vasconcelos~\cite{DBLP:journals/jfp/GayV10}.

The description so far produces a \emph{linear} session typed language. In
particular we cannot take advantage of recursive types for there is no support
for consuming such a type. Recursive functions are the usual means of consuming
recursive types, a \lstinline|Stream| for example, but they constitute the
finished example of non-linear resources. A simple way out is to annotate
entries in the typing context with the number of times a resource can be used,
along the lines of Linear Haskell~\cite{DBLP:journals/pacmpl/BernardyBNJS18} or
Quantitative Type
Theory~\cite{DBLP:conf/lics/Atkey18,DBLP:conf/birthday/McBride16} (also used in
Nominal Session Types~\cite{DBLP:journals/pacmpl/MordidoS0V23}).
An alternative would be to introduce shared resources, functions in particular,
in the linear type system~\cite{DBLP:journals/jfp/GayV10}.

The $\keyword{fork}$ operator creates a new thread. Yet the term language
depicted so far features no support for threads running concurrently. This is
usually achieved by introducing a separate language for \emph{processes},
denoted by $P,Q$. The basic processes are terms, for example of the form
$\langle M\rangle$, composed by means of the parallel composition of two
processes, $P\mid Q$, and of scope restriction, $(\nu xy)P$, describing the two end
points $x$ and $y$ of a channel circumscribed to process $P$. Processes come
equipped with a notion of reduction, featuring axioms for output-input,
select-branch, and close-wait interactions, complemented with suitable
congruence rules. For example, the axiom, for output-input might be as follows,
\begin{equation*}
  (\nu xy)(\langle E[\send\,v\,x]\rangle \mid \langle F[\receive\,y]\rangle)
  \rightarrow
  (\nu xy)(\langle E[x]\rangle \mid \langle F[(v, y)]\rangle)
\end{equation*}
where the $\send\,v\,x$ term in evaluation context $E$ is rewritten in channel
$x$ (so that term $E[x]$ may continue interaction on $x$) and the $\receive\,y$
term in evaluation context $F$ is rewritten in a pair featuring the value
received $v$ and the continuation channel $y$ (so that term $F[(v, y)]$ may use
the value $v$ in the message while continuing the interaction on $y$). Details
can be found in different
sources~\cite{DBLP:journals/iandc/AlmeidaMTV22,DBLP:journals/jfp/GayV10}.

\section{Conclusion and future work}
\label{section:conclusion}

We show how to integrate staged metaprogramming into a linear lambda calculus
and sketch how to extend the language to concurrency and session types.
The type system we propose is deliberately non-algorithmic. We believe that
standard techniques---including an explicitly typed and level-annotated
syntax~\cite{DBLP:journals/pacmpl/JangGMP22} and having the typing rules
``return'' the unused part of the
context~\cite{walker:substructural-type-systems}---would lead to algorithmic
type checking.



\paragraph{Acknowledgements}

\begin{sloppypar}
  This work was partly supported by JSPS Invitational Short-Term
  Fellowships for Research in Japan.
  It was further supported by the FCT through
  project SafeSessions
  (doi: \href{https://doi.org/10.54499/PTDC/CCI-COM/6453/2020}{10.54499/PTDC/CCI-COM/6453/2020}),
  by the LASIGE Research Unit
  (doi: \href{https://doi.org/10.54499/UIDB/00408/2020}{10.54499/UIDB/00408/2020} and
   doi: \href{https://doi.org/10.54499/UIDP/00408/2020}{10.54499/UIDP/00408/2020}),
  and by the LIACC Research Unit
  (doi: \href{https://doi.org/10.54499/UIDB/00027/2020}{10.54499/UIDB/00027/2020} and
   doi: \href{https://doi.org/10.54499/UIDP/00027/2020}{10.54499/UIDP/00027/2020}),
\end{sloppypar}



\bibliographystyle{eptcs}
\bibliography{biblio.bib} 

\newpage

\appendix

\end{document}